\documentclass[12pt]{iopart}
\usepackage{epsfig}
\usepackage{iopams}

\newcommand{\be}{\begin{equation}}
\newcommand{\ee}{\end{equation}}

\begin{document}

\title[Constructive Quantum Shannon Decomposition from Cartan Involutions]{Constructive Quantum Shannon Decomposition from Cartan Involutions}

\author{Byron Drury, Peter Love}

\address{Department of Physics, 370 Lancaster Ave., Haverford College,
Haverford, PA U.S.A. 19041, USA}
\ead{plove@haverford.edu}
\begin{abstract}
The work presented here extends upon the best known universal quantum circuit, the Quantum Shannon Decomposition proposed in [Vivek V. Shende, Stephen S. Bullock and Igor Markov, {\em Synthesis of Quantum Logic Circuits}, {\em IEEE Trans. Comput.-Aided Des. Integr. Circuits Syst. 25 (6): 1000-1010 (2006)}]. We obtain the basis of the circuit's design in a pair of Cartan decompositions. This insight gives a simple constructive algorithm for obtaining the Quantum Shannon Decomposition of a given unitary matrix in terms of the corresponding Cartan involutions.
\end{abstract}
\vspace{2pc}

\maketitle

\section{Introduction}
Quantum computation has attracted interest in recent years because it appears to violate the strong form of the Church-Turing thesis; quantum computers seem to be fundamentally more powerful than any possible classical computer~\cite{Deutsch85}.  In 1994 Peter Shor published efficient quantum algorithms for the prime factorization of integers and the calculation of discrete logarithms modulo arbitrary primes~\cite{shor-1997-26}.  Lov Grover's 1995 introduction of the quantum search algorithm provided a polynomial speedup for unstructured searches~\cite{PhysRevLett.79.325}.  As early as 1982 Richard Feynman pointed out the inherent difficulties in simulating quantum systems with classical processors and suggested the possibility that the use of quantum information processing could produce exponential speedups in such simulations~\cite{1982IJTP...21..467F}.  Subsequently efficient quantum algorithms for performing simulations of physical systems were developed~\cite{wiesner-1996,1998,simulators, PhysRevLett.79.2586, quantsystsim,plove,kassal-2008}, vindicating Feynman's prediction and further motivating theoretical and experimental work towards realizing quantum computation.  

In this paper we focus on the quantum circuit model of quantum computation~\cite{yao1993}. In this setting a quantum computation is a unitary transformation applied to $n$ ideal qubits (we ignore decoherence throughout). Given the irrelevance of global phases the set of all such transformations is the special unitary group $SU(2^n )$. To represent an element of $SU(2^n)$ by a circuit we must specify a fixed set of elementary gates which act on a fixed number of qubits. A typical choice is the controlled-NOT (CNOT) and arbitrary one-qubit gates. The length of a circuit is the number of elementary gates which it contains, however, because of the relative difficulty of multi-qubit operations we shall only consider the number of CNOT gates in a circuit. There are several means of physically implementing a quantum computation~\cite{NeilA.Gershenfeld01171997,2003Natur.421...48G,PhysRevB.70.184525,2001Natur.414..883V,2001OptSp..91..429W,jones:1648}. One qubit local operations and a few two qubit operations, such as the controlled-NOT (CNOT) gate have been experimentally implemented~\cite{544199,phase,PhysRevLett.75.4714,PhysRevA.68.032316,zhao:030501,PhysRevLett.83.5166}.   

The set of all allowed transformations for a quantum computer form the group $SU(2^n)$ and a generic element of $SU(2^n)$ requires a circuit of length $\mathcal{O}(4^n)$ gates.  Specific transformations corresponding to efficient quantum algorithms are of particular interest. A quantum algorithm specifies a circuit family, with a circuit defined for each value of $n$. For a quantum algorithm to be efficient each of these circuits must be composed of a number of operations bounded above by a polynomial in $n$. Each of these operations must involve a subset of the $n$ qubits with size bounded above by a polynomial in the logarithm of $n$. Some algorithms, for example the quantum Fourier transform, naturally decompose into elementary gates acting on qubits~\cite{Josza1998}. In other cases, for example generic quantum Fourier transforms of functions on groups other than $Z/({2^n}Z)$~\cite{beals1997,cmoore2006}, and in application of phase estimation to problems of quantum simulation~\cite{plove,PhysRevLett.79.2586}, bounded size operations arise which do not naturally factor into elementary gates. Before such quantum algorithms may be implemented experimentally one is therefore faced with a problem of quantum compilation - given a set of unitary operators of fixed size and an elementary gate set, constructively produce the quantum circuit realizing the operators. 

It was shown by construction in 1995 that the set of one qubit operations and the CNOT are universal: any unitary operation on any number of qubits can be realized as a circuit over these gates. However, the number of CNOT gates required for $n$ qubits was of order $n^3 4^n$~\cite{PhysRevA.52.3457}. Since 1995 a number of advances have been made towards the CNOT optimization of universal quantum circuits.  We divide these into three categories: circuit optimization, Lie algebra decompositions, and explicit algorithms.  

Knill proved that the asymptotic CNOT cost of universal quantum circuits could be reduced by a factor of $n^2$ to $O(n 4^n)$~\cite{ knill95approximation}.  In 2004 Shende, Markov and Bullock proved the highest known lower bound on asymptotic CNOT cost, $\lceil\frac{1}{4}(4^n -3n -1) \rceil$~\cite{PhysRevA.69.062321}, and Vartiainen, M\"{o}tt\"{o}nen, and Salomaa simplified the best existing circuit using Gray codes to achieve for the first time a leading order CNOT cost of $O(4^n)$ (in fact, for large $n$, the cost was approximately $8.7\times 4^n$), a multiplicative factor away from the highest known lower bound~\cite{PhysRevLett.92.177902}.  Later that year, the same authors, along with Bergholm, presented a decomposition based on the cosine-sine matrix decomposition (CSD) which produced asymptotic behavior scaling as $4^n - 2^{n+1}$~\cite{PhysRevLett.93.130502}.  Vatan and Williams published a three CNOT universal two qubit gate along with a proof that fewer CNOTs could never achieve universality~\cite{vatan:032315}, and proposed a 40 CNOT universal three qubit gate which was, at the time, the best known~\cite{vatan-2004}. The current best known circuit decomposition applicable to systems of more than two qubits was introduced by Shende, Bullock and Markov. Using intuition drawn from the Shannon decomposition of classical logic circuit design, along with the application of some circuit identities, Shende, Bullock and Markov have designed a universal circuit requiring 20 CNOTs in the three qubit case and $\frac{23}{48}4^n-\frac{3}{2}2^n+\frac{4}{3}$ CNOTs asymptotically~\cite{Synthesis}. This decomposition is known as the Quantum Shannon Decomposition (QSD), by analogy with the Shannon decomposition of classical circuit design, and brings the upper bound on asymptotic CNOT cost to within a factor of two of the highest known lower bound while halving the cost of implementing a general three qubit gate to $20$ CNOTs.   

The second area of research is the exploration of the various ways of decomposing the Lie algebra of the special unitary group.  Essentially all of the work in this area has made use of the Cartan decomposition.  In the first part of the twentieth century Cartan proved that (up to conjugacy) there exist only three types of Cartan decomposition on the unitary lie algebra, \textbf{AI-III}~\cite{Cartan1, Cartan2}. The CNOT optimal two qubit circuit of Vatan and Williams~\cite{vatan:032315} is, as described in detail below, based on a type \textbf{AI} Cartan decomposition. Khaneja and Glaser proposed a scheme based on a Cartan decomposition of $\mathfrak{su}(2^n)$ (now known as the Khaneja Glaser Decomposition, or KGD) which lends itself to efficient recursive circuit decompositions~\cite{Khaneja2001}, and, working with Brockett, they showed that this scheme was time optimal for NMR based implementations of quantum computation~\cite{PhysRevA.63.032308}. Bullock identified the Khaneja Glaser Decomposition, as well as the CSD, as type \textbf{AIII} Cartan decompositions and thereby established an equivalence between the two~\cite{bullock-2004-4}.   The KGD was used by Vatan and Williams to produce their efficient two and three qubit circuits~\cite{vatan-2004, vatan:032315}.  Bullock and Brennen and more recently Dagli, D'Alessandro and Smith have used type \textbf{AI} and \textbf{AII} decompositions, including the Concurrence Canonical Decomposition (CCD) and the Odd-Even Decomposition (OED),  to study entanglement dynamics in quantum circuits~\cite{bullock:2447,dagli}.   

In order to make practical use of a CNOT optimized quantum circuit or a novel Lie algebra decomposition it is necessary to have an algorithm which can extract the parameters which appear in the decomposition from an arbitrary unitary operation.  Sousa and Ramos provided an algorithm based on the generalized singular value decomposition for computing the parameters in a CNOT optimized two qubit circuit (the parameters for Vatan and Williams circuit can be extracted from their algorithm with a little algebra, and other equivalent circuits can be computed with a similar amount effort)~\cite{sousa-2006}.  Just as Vatan and Williams' work on small numbers of qubits does not generalize to larger operators, however, Sousa and Ramos' algorithm does not generalize beyond two qubits. Earp and Pachos provided a constructive algorithm to perform a type \textbf{AIII} Cartan decomposition of an arbitrary $n$ qubit operator (they use the Khaneja Glaser Decomposition specifically, but their algorithm can be modified to implement other forms of the \textbf{AIII} decomposition)~\cite{earp:082108}.  Earp and Pachos' algorithm relies on numerical optimization and a truncation of the Baker-Campbell-Hausdorff formula. Nakajima, Kawano and Sekigawa published the first algorithm to compute Cartan decompositions of the unitary group making explicit use of Cartan involutions~\cite{nakajima-2005}; their algorithm computes parameters for circuits composed of uniformly controlled operations, similar to the circuits produced by CSD based schemes. Their algorithm requires $4^n-2^{n-1}$ CNOT gates asymptotically. In the three-qubit case this number can be reduced by taking advantage of the known CNOT-optimized two qubit circuit developed by Vatan and Williams to produce a 44 CNOT universal three qubit circuit (see Fig. ~\ref{44cnot}).  Since a lower bound of $\frac{1}{4}(4^n-3n-1)$ has been proven on the asymptotic CNOT cost of arbitrary n-qubit operations with a lower bound of 14 CNOTs in the three qubit case ~\cite{PhysRevA.69.062321}, this efficiency cannot be improved by more than a factor of four. Circuits produced by Nakajima, Kawano and Sekigawa's algorithm are a factor of two longer than circuits obtained from the Quantum Shannon Decomposition (QSD). However, the QSD lacks a constructive Lie algebra based factoring algorithm in the published literature so far.  It is to this issue we turn in the remainder of the paper.  

We first give some mathematical background introducing important definitions and theorems which will be used later in the work.  We then discuss the important special cases of one and two qubit operations, and provide Cartan involution based algorithms for extracting parameters for CNOT optimal quantum circuits from arbitrary one and two qubit unitary operations. We then place the QSD, the best known circuit decomposition in terms of CNOT cost, into a Lie algebraic context by showing it to be an alternating series of Cartan decompositions.  We define the Cartan involutions which correspond to these decompositions, and we show that these involutions can be used recursively to obtain the QSD for unitary operators on any number of qubits.

\section{Mathematical Background}

In the interest of making our presentation more self-contained, we briefly review some basic definitions which will be important throughout this work. For a fuller presentation we refer the reader to~\cite{Carter,helgason}. Throughout, we use $[ab]$ to denote the Lie bracket in general, and the notation $[a,b]$ to denote the Lie bracket for matrix algebras where it is the commutator $[a,b] = ab-ba$.

\subparagraph{Definition 1:} If a subalgebra $I$ of a Lie algebra $\mathfrak{g}$ satisfies the condition that $ [xy]\in I$ for all $x\in \mathfrak{g},\: y\in I $ then $I$ is called an \textit{ideal} in $\mathfrak{g}$.  

\subparagraph{Example 1:} Clearly $0$ and $\mathfrak{g}$ are trivial ideals of $\mathfrak{g}$.  An important example of an ideal is the \textit{derived algebra} of $\mathfrak{g}$, denoted $[\mathfrak{gg}]$, which consists of all linear combinations of brackets $[xy]$, with $x,y \in \mathfrak{g}$.  

\subparagraph{Definition 2:} A non-abelian Lie algebra $\mathfrak{U}$ (i.e. $[\mathfrak{UU}]\neq 0$) in which the only ideals are $0$ and all of $\mathfrak{U}$ is called \textit{simple}.  Observe that since the derived algebra is an ideal, for any simple Lie algebra $S$ the derived algebra is equal to the entire algebra: $[SS]=S$.

We may define a sequence of ideals, the \textit{derived series} of an algebra $A$, as follows: \[A^{(0)} =A,\: A^{(1)} =[AA],\: A^{(2)} =[A^{(1)} A^{(1)} ],\: ...,\: A^{(i)} =[A^{(i-1)} A^{(i-1)} ],\: ... \] If $A^{(n)}=0$ for some $n$ we call $A$ \textit{solvable}.  Observe that all abelian Lie algebras are solvable, while all simple Lie algebras are nonsolvable.  We shall simply state the fact that every Lie algebra contains a unique maximal solvable ideal (maximal in the sense that it is contained in no larger solvable ideal), which is referred to as the \textit{radical} of the algebra.  If $L$ is a non-zero Lie algebra and Rad $L=0$, we call $L$ \textit{semi-simple}.  This condition for the semi-simplicity of a Lie algebra is equivalent to the condition that the algebra is the direct sum of simple Lie algebras.  Most of the Lie algebras which occur in physics are semi-simple, and there exists a very rich and well developed structure theory of semi-simple Lie algebras which we shall exploit throughout the remainder of this work.  The essential structure theorem which lies behind both the CSD, the KGD, and as we shall show later the QSD, is the Cartan decomposition.  

\subparagraph{Definition 3:}A \textit{Cartan Decomposition} of a real semi-simple Lie algebra $\mathfrak{g}$ is a decomposition $\mathfrak{g=m\oplus k}$ where $\mathfrak{m=k^{\bot}}$, for which $\mathfrak{k}$ and $\mathfrak{m}$ satisfy the commutation relations: 
\be \left[\mathfrak{k},\mathfrak{k}\right]\subset\mathfrak{k}\ee
\be \left[\mathfrak{m},\mathfrak{k}\right]=\mathfrak{m}\ee
\be \left[\mathfrak{m},\mathfrak{m}\right]\subset\mathfrak{k}\ee

A few further features of the Cartan decomposition are essential.  

\subparagraph{Definition 4:} Consider a semi-simple Lie algebra with Cartan decomposition $\mathfrak{g=m\oplus k}$ and a subalgebra $\mathfrak{h}$ of $\mathfrak{g}$ contained in $\mathfrak{m}$.  Because $\left[\mathfrak{m},\mathfrak{m}\right]\subset\mathfrak{k}$, $\mathfrak{h}$ must be Abelian.  We refer to a maximal Abelian subalgebra contained in $\mathfrak{m}$ as a \textit{Cartan subalgebra} of $\mathfrak{g}$ and $\mathfrak{k}$.  

\subparagraph{Definition 5:} The Lie group $G$ acts on its Lie algebra $\mathfrak{g}$ through a conjugation, known as the adjoint action, $Ad_{G}:\:\mathfrak{g\rightarrow g}$ defined by
\be Ad_{U}X=U^\dagger XU  \ee
for $u\in G$ and $X\in\mathfrak{g}$, and for $K=\exp(\mathfrak{k})$ we define the \textit{Adjoint orbit} of $X$ to be 
\be Ad_{K}X=\bigcup_{k\in K}Ad_{k}X\ee

Any two Cartan subalgebras $\mathfrak{h}$ and $\mathfrak{h'}$ are related to one another through the adjoint action of the group $G$ on its Lie algebra $\mathfrak{g}$.  With these definitions, we now state

\subparagraph{Theorem 1:} For any two maximal Abelian subalgebras $\mathfrak{h}$ and $\mathfrak{h'}$ in $\mathfrak{m}$ there is an element $k\in K$ such that $Ad_{k}(\mathfrak{h})=\mathfrak{h'}$.  Furthermore, the adjoint orbit of $\mathfrak{h}$ is equal to $\mathfrak{m}$, \textit{i.e.}  \be \mathfrak{m} = \bigcup_{k\in K} Ad_k \mathfrak{h}\ee

Finally, we come to the key definition in this paper:

\subparagraph{Definition 6:} Given a semisimple Lie algebra $\mathfrak{g}$ with Cartan decomposition $\mathfrak{g=m\oplus k}$ and a Cartan subalgebra $\mathfrak{h}$, let $A=\exp(\mathfrak{h})$ and $K=\exp(\mathfrak{k})$, then $G=KAK$ is called a \textit{(global) Cartan decomposition} of the semi-simple Lie group $G$.  

The theorem which establishes the existence of such a decomposition for any semi-simple Lie group is proved in ~\cite{helgason,Carter,gilmore}.  The $G=KAK$ structure has been used widely in work on quantum circuit decompositions in the past, most notably in Khaneja and Glaser's work, as well as in CSD based circuit designs (as explained by Bullock~\cite{bullock-2004-4}) and in subsequent work based on these decompositions (cf. e.g.~\cite{Khaneja2001,PhysRevLett.93.130502,vatan-2004,dagli}).  The task of computing the Cartan factors for a specific unitary matrix is greatly facilitated by the existence of Cartan involutions.

\subparagraph{Definition 7:} A Cartan involution, denoted $\theta$, is a non-identity automorphism on a Lie algebra $\mathfrak{u}$ such that $\theta^{2}$ is the identity, and the global Cartan involution has the equivalent action on $U = \exp (\mathfrak{u})$ with the property that 
\be \label{involution}\theta (g)=\left\{\begin{array}{cl}g & \:\:g\in\mathfrak{k}\\ -g & \:\:g\in\mathfrak{m}\end{array}\right.,\:\:\:\:\:\:\:\Theta (G)=\left\{\begin{array}{cl}G & \:\:G\in\exp (\mathfrak{k})\\ G^\dagger & \:\:G\in\exp (\mathfrak{m})\end{array}\right.\ee 

In the case of $\mathfrak{su}(n)$ there are only three classes of Cartan decomposition, denoted \textbf{AI}, \textbf{AII}, and \textbf{AIII}.  The $\mathfrak{k}$ subalgebras of $\mathfrak{su}(n)$ are isomorphic to $\mathfrak{so}(n)$, $\mathfrak{sp}(\frac{n}{2})$, and $\mathfrak{s[u}(p)\oplus \mathfrak{u}(q)]$ for any $p+q = n$ for \textbf{AI, AII}, and \textbf{AIII} decompositions, respectively (\textbf{AII} only exists for unitary groups acting on an even number of dimensions, a common situation in quantum information where the state-spaces of $n$-qubit registers have dimension $2^n$)~\cite{dagli}.  In this work we are particularly concerned with decompositions of type \textbf{AI} and \textbf{AIII} because in certain important cases there are straightforward and efficient means of physically implementing real orthogonal or direct sum unitary operators.  Since we are concerned in this work only with the unitary group, whose elements satisfy the condition $U^{-1} = U^{\dagger}$, we may exploit the Cartan involution to factor matrices.  

\subparagraph{Theorem 2:} For any $G\in SU(2^{n})$ with Cartan decomposition $G=KM$, $K\in \exp(\mathfrak{k})$, $M\in \exp(\mathfrak{m})$, $M^{2}$ is uniquely determined by $M^{2}=\Theta(G^{\dagger})G$.

\subparagraph{Proof:} $\Theta(G^{\dagger})G=\Theta(M^{\dagger}K^{\dagger})KM=\Theta(M^{\dagger})\Theta(K^{\dagger})KM=MK^{\dagger}KM=M^{2}$. $\Box$

\bigskip

A $KAK$ type decomposition of the special unitary group is desirable because there is a considerable amount of freedom in selecting the $\mathfrak{k}$ subalgebra and a Cartan subalgebra $\mathfrak{h}$, and with appropriate selection of $\mathfrak{k}$ and $\mathfrak{h}$ the factors returned for an arbitrary unitary operator are of a form which may readily be translated into physically realizable quantum gate sequences.  Indeed, the Khaneja-Glaser Decomposition has been shown to be time optimal for NMR quantum computing, as compared to other published decompositions~\cite{PhysRevA.63.032308}.  The existence of this decomposition is of no practical use, however, without an algorithm for explicitly calculating the factors $K_{1}$, $K_{2}$ and $A$ for a given specific unitary matrix.  

\subparagraph{Notation} When discussing the generators of the Lie algebras of multi-qubit operator groups we will use a streamlined notation.  We define $ZI=\sigma_{z}\otimes\textbf{1}$, $IX=\textbf{1}\otimes\sigma_{x}$, $ZY=\sigma_{z} \otimes \sigma_{y}$ and so on, where $ \sigma_{x},\: \sigma_{y}$ and $ \sigma_{z}$ are the familiar Pauli spin matrices \[ \sigma_{x} = \left(\begin{array}{cc}0 & 1 \\1 & 0\end{array}\right),\:\:\:  \sigma_{y} = \left(\begin{array}{cc}0 & -i \\i & 0\end{array}\right)\:\:\:,  \sigma_{z} = \left(\begin{array}{cc}1 & 0 \\0 & -1\end{array}\right).   \]  Additionally, we define $X^{(n)}$ to be a Pauli-$x$ (likewise $y$ and $z$) acting on the $n^{th}$ qubit, i.e. $Z^{(3)} = IIZ$.  

\section{Special Cases: One and Two Qubits}
\subsection{One qubit factoring: Euler Angle decomposition of $SU(2)$ as a Cartan Decomposition}
We now provide a simple, illustrative example of a Cartan decomposition and an involution based algorithm for converting an arbitrary one qubit unitary operator into a Cartan inspired circuit.  This is the simplest possible case of a Cartan decomposition of a unitary group, however, the factoring of multi-qubit gates inevitably reduces in the end to a series of one-qubit gates which must themselves be decomposed. The structure of the algorithm for this simple example is identical to the more involved cases to follow.

\subparagraph{Definition:} The Lie algebra $\mathfrak{su}(2)$ is generated by the Pauli spin matrices.  The decomposition $\mathfrak{su}(2)=\mathfrak{k}\oplus \mathfrak{m}$ where $\mathfrak{k}= {\rm span}_{\mathbb{R}}i\{Y\}$ and $\mathfrak{m}={\rm span}_{\mathbb{R}}i\{X,Z\}$ satisfies the criteria to be a Cartan decomposition.  Furthermore, either $ {\rm span}_{\mathbb{R}}i\{X\}$ or $ {\rm span} _{\mathbb{R}}i\{Z\}$ is a maximally abelian subalgebra of $\mathfrak{su}(2)$ contained in $\mathfrak{m}$.  Thus the decomposition of $SU(2)$ given by $U=e^{iAY}e^{iBZ}e^{iCY}$ is a Cartan decomposition. Using the fact that $SU(2)$ is the double cover of $SO(3)$, we recognize this Cartan decomposition as the Euler angle decomposition of three dimensional rotations.  We now explicitly calculate the Euler angle decomposition of an arbitrary single qubit unitary using a Cartan involution.

The Cartan involution corresponding to our chosen Cartan decomposition ($\mathfrak{k}= {\rm span}_{\mathbb{R}}i\{Y\}$, $\mathfrak{m}={\rm span}_{\mathbb{R}}i\{X,Z\}$ and $\mathfrak{h} = {\rm span} _{\mathbb{R}}i\{Z\}$) is $\theta (u) = Y u Y,\;\Theta (U) = Y U Y$.  We compute the Cartan $KAK$ decomposition of an arbitrary $G \in SU(2)$ as follows
\paragraph{1.} We exploit Theorem 2 to calculate $M^2 = Y G^{\dagger} Y G$
\paragraph{2.} Diagonalize $M^2 = PDP^{\dagger}$.  Note that as a diagonal element of $SU(2)$, $D$ must be of the form $e^{i\alpha Z}$, i.e. $D\in \exp (\mathfrak{h})$, and, furthermore, Theorem 1 implies that $P\in \exp (\mathfrak{k})$.
\paragraph{3.} We now have $M = PD^{1/2}P^{\dagger}$ and we may find $K = GM^{\dagger}$.  
\paragraph{4.} This constitutes a complete decomposition of $G$ into the form $e^{iA Y}e^{iB Z}e^{iC Y}$: $G = KPD^{1/2}P^{\dagger}$, and it is trivial to extract the angles $A$, $B$ and $C$ from the matrix forms of these operators.  

\subsection{Two qubit factoring from a Cartan decomposition.}

\begin{figure}
\begin{center}
\includegraphics[width=0.8\textwidth,viewport=0 460 400 600,clip]{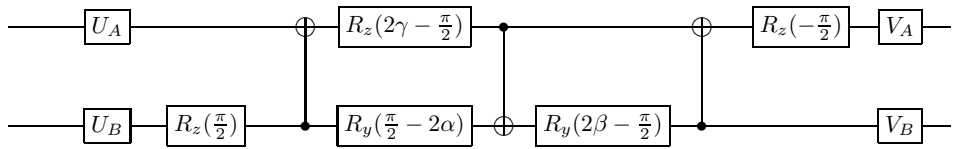}
\caption{\label{2q}The CNOT optimized universal two qubit circuit; $U_{A},\,U_{B},\,V_{A},\,{\rm and}\,V_{B}$ may be decomposed into 3 single qubit rotations each by the Euler angle decomposition given above, and $V_A$ and $U_B$ may absorb the z-rotations preceding and following them respectively yielding a circuit consisting of 3 CNOT gates and 15 single qubit rotations.}
\end{center}
\end{figure}

\begin{figure}
\begin{center}
\includegraphics[width=1.0\textwidth,viewport=0 150 500 600,clip]{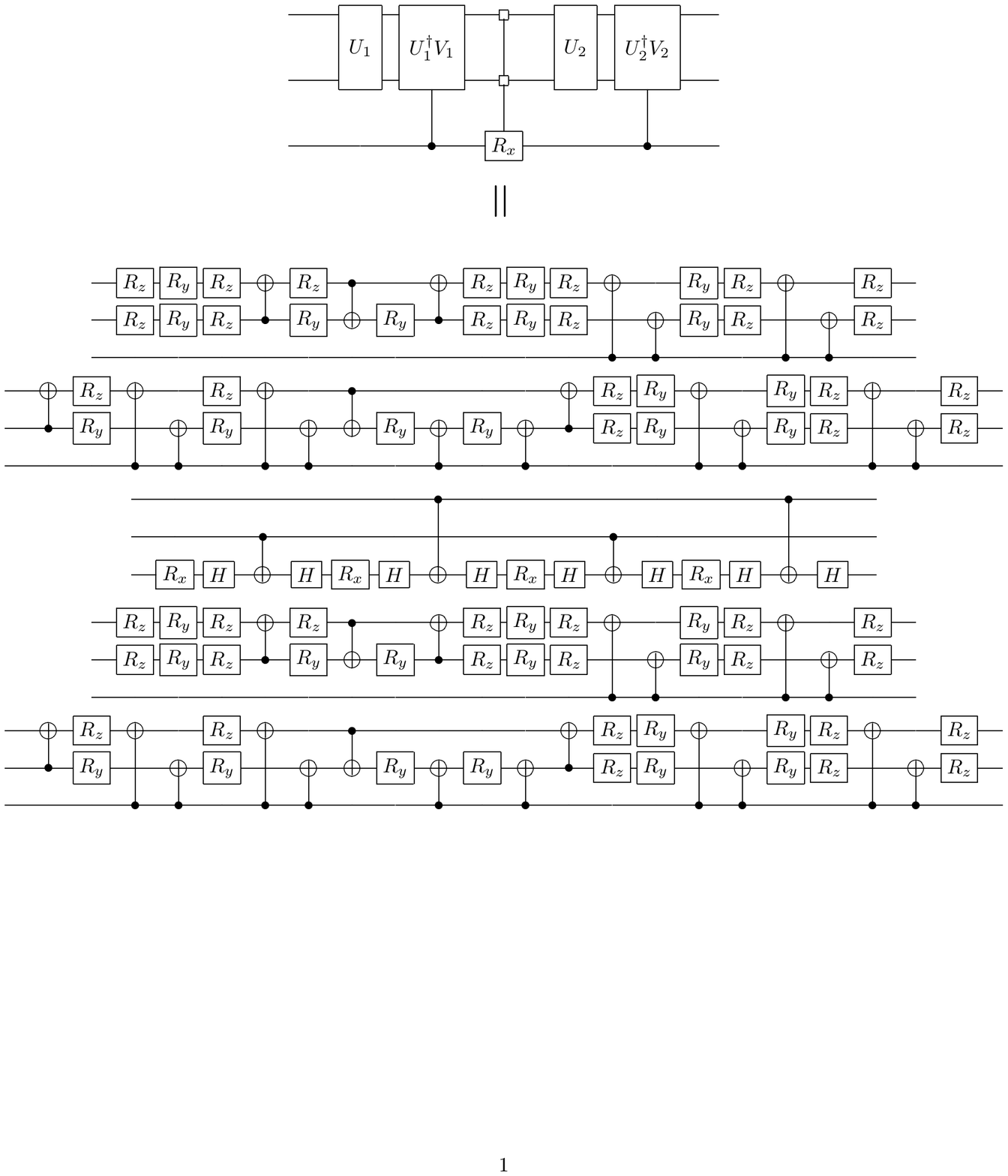}
\caption{\label{44cnot}A simplified  three qubit circuit based on Nakajima, Kawano and Sekigawa's algorithm: uniformly controlled two qubit operations are built using Vatan and Williams' optimal two qubit circuit to produce a universal 44 CNOT three qubit circuit with a constructive algorithm.  The operator represented here is $(U_{1}\otimes|0\rangle\langle 0|+V_{1}\otimes|1\rangle\langle 1|)(R_{x1} \oplus R_{x2} \oplus R_{x3} \oplus R_{x4})(U_{2}\otimes|0\rangle\langle 0|+V_{2}\otimes|1\rangle\langle 1|)$, in accordance with the NKS algorithm.}
\end{center}
\end{figure}

The task of factoring two qubit operators is facilitated by several unique properties of $SU(4)$. Firstly, $SO(4)$ is the Lie group corresponding to the $\mathfrak{k}$ subalgebra of $\mathfrak{su}(4)$ under a type \textbf{AI} involution.  $SO(4)$ and the group of local operations acting on two qubits separately, $SU(2)\otimes SU(2)$, share a simply connected covering group, $Spin(4)$.  In fact, elements of $SO(4)$ are mapped uniquely onto elements of $SU(2)\otimes SU(2)$ by changing to the ``magic basis" of Bell states through conjugation by the matrix~\cite{vatan:032315}:  
\be  B=\frac{1}{\sqrt{2}}\left(\begin{array}{cccc}1 & i & 0 & 0 \\0 & 0 & i &
1 \\0 & 0 & i & -1 \\1 & -i & 0 & 0\end{array}\right).\label{bell}\ee   
There is no equivalent connection between $SO(2^n )$ and $SU(2^{n-1})\otimes SU(2^{n-1})$ for $n>2$.  As a result of this close connection between the type \textbf{AI} Cartan decomposition of $SU(4)$ and the group of local operations (which may be implemented without the use of CNOT gates) it is possible to construct a universal 2 qubit circuit requiring only 3 CNOT gates in the worst case (see Figure~\ref{2q})~\cite{vatan:032315,PhysRevLett.91.027903,sousa-2006,shende:012310}. 

\subparagraph{Definition:} The involution for type \textbf{AI} Cartan decompositions of $\mathfrak{su}(N)$ is given by \be \theta(u) = -u^T \:\: {\rm for} \: u\in\mathfrak{su}(N),\:\:\:\Theta(U) = (U^{-1})^T = U^* \:\: {\rm for} \: U\in SU(N). \label{ai}\ee

The involution given by~(\ref{ai}) fixes a $\mathfrak{k}$-subalgebra corresponding to $\mathfrak{so}(4)$, $\mathfrak{k} = {\rm span}_{\mathbb{R}}i\{IY,XY,ZY,YI,YX,YX\}$, and the diagonal elements of $\mathfrak{m}$, i.e. $ \mathfrak{h} = {\rm span}_{\mathbb{R}}i\{IZ,ZI,ZZ\}$ constitute a Cartan subalgebra.  Furthermore, as discussed in the introduction, a transformation to the basis of Bell states (the ``magic basis") maps this $\mathfrak{k}$ subalgebra onto $\mathfrak{su}(2) \oplus \mathfrak{su}(2)$ and also maps the maximal abelian subalgebra of diagonal matrices onto the subalgebra chosen by both Khaneja and Glaser and Vatan and Williams, $\mathfrak{h'} = {\rm span}_{\mathbb{R}}i\{XX,YY,ZZ\}$.  As a result, we may use the Cartan involution of Equation~(\ref{ai}) and matrix diagonalization to compute the parameters necessary for Vatan and Williams two-qubit CNOT optimal circuit.  

The parameters for an arbitrary two qubit unitary $U$ may be calculated as follows:
\paragraph{1.} We define a new operator $U'= B^{\dagger} U B$ where $B$ is defined in Equation~\ref{bell}.
\paragraph{2.} Compute $M^2 = \Theta (U'^{\dagger})U' = (U'^{\dagger})^* U' = U'^T U'$, which is in the exponentiation of $\mathfrak{m}$.
\paragraph{3.} Diagonalize: $M^2 = P D P^{\dagger}$ where $D \in \exp(\mathfrak{h})$ and $P \in SO(4)$.
\paragraph{4.} Find $D^{\frac{1}{2}}$ and hence $K' = U' P D^{-\frac{1}{2}} P^{\dagger}$.
\paragraph{5.} $K'P$ and $P^{\dagger}$ are both elements of $SO(4)$, so $K_1 = BK'PB^{\dagger}$ and $K_2 = BP^{\dagger}B^{\dagger} \in SU(2)\otimes SU(2)$ and $A = BD^{\frac{1}{2}} B^{\dagger}\in \exp(\mathfrak{h'})$.  Hence \be K_1 A K_2 = BK'PB^{\dagger}BD^{\frac{1}{2}} B^{\dagger}BP^{\dagger}B^{\dagger} = BK'PD^{\frac{1}{2}}P^{\dagger}B^{\dagger} = B U' B^{\dagger} = U \ee is a Cartan decomposition of $U$ of the type used by Vatan and Williams. 
\paragraph{6.} Simple algebraic manipulations of $D^{\frac{1}{2}}$ yield the parameters $\alpha, \: \beta \:{\rm and }\: \gamma$ which appear in the center portion of the circuit in Figure 6 of~\cite{vatan:032315} and the partial trace may be used to separate $K_1$ and $K_2$ into the local operations of which they are composed, which may then be decomposed as described in the previous section.  

\section{The QSD from Cartan Involutions}

In this Section we give a Cartan decomposition and constructive algorithm for obtaining the QSD (recall that it is not possible to exceed the QSD's efficiency by even a factor of two for any number of qubits). This algorithm is constructive and produces circuits which are less than half as long as the constructive algorithms of Nakajima, Kawano and Sekigawa~\cite{earp:082108,nakajima-2005}. The principle difference between those algorithms and the QSD is that they proceed by reducing an $n$-qubit circuit to a circuit involving uniformly controlled $n-1$ qubit gates. These uniformly controlled gates are then reduced to controlled and uncontrolled $n-1$ qubit gates. The uncontrolled $n-1$ qubit gates, and the controlled $n-1$ qubit gates are then factored again using the Cartan decomposition. However, all gates obtained by this decomposition must be controlled, leading to a doubling of the number of CNOTs over the best known decompositions. This problem arises because only part of the decomposition is handled at the Lie algebra level - after the first decomposition circuit identities are introduced before the Cartan decomposition is applied again. In what follows we take the Lie algebraic point of view throughout: the uniformly controlled operations are treated as a Lie-subgroup, and a Cartan decomposition of the corresponding Lie-subalgebra is obtained. This Cartan decomposition results in uncontrolled $n-1$ qubit operations which remain to be factored, and so the first part of the algorithm of Nakajima, Kawano and Sekigawa can be applied again. The resulting algorithm is an alternating pair of Cartan decompositions, each of which has a simple Cartan involution which enables the factors to be obtained explicitly. Inspection of the resulting procedure reveals precisely the QSD of~\cite{Synthesis} and so this algorithm gives a Cartan decomposition based derivation of the QSD and a Cartan involution based explicit algorithm for obtaining the QSD. 

Because every other step in our recursive procedure is identical to the first step of Nakajima, Kawano and Sekigawa's algorithm, we first define the correponding components $\mathfrak{k}$ and $\mathfrak{m}$ of the Cartan decomposition of $SU(2^{n})$, and the Cartan subalgebra $\mathfrak{h}$.  The $\mathfrak{k}$-subalgebra is of type \textbf{AIII}: the direct sum of two lower dimensional unitary Lie algebras $\mathfrak{k=s[u}(p)\oplus \mathfrak{u}(q)]$ where $p+q=2^{n}$.  
\subparagraph{Definition:} For the $n$-qubit case the decomposition is defined by: 
\be \mathfrak{k}={\rm span}_\mathbb{R}\{A\otimes Z,B\otimes{\bf 1}, i Z^{(n)}|A,B\in\mathfrak{su}(2^{n-1})\}\ee
\be \mathfrak{m}={\rm span}_\mathbb{R}\{A\otimes X,B\otimes Y, i X^{(n)},  i Y^{(n)}|A,B\in\mathfrak{su}(2^{n-1})\}\ee
\subparagraph{Definition:} The Cartan involution is:
\be \theta (u)=Z^{(n)}uZ^{(n)},\:\:\:\:\:\: \Theta (U)=Z^{(n)}UZ^{(n)}\ee
Hence we may compute the global Cartan decomposition $G=KM$ of $SU(2^n )$ as in Theorem 2.  

We must now define a Cartan subalgebra $\mathfrak{h}$ contained in $\mathfrak{m}$.  Here Nakajima et al. make a different choice of $\mathfrak{h}$ to that used by Khaneja and Glaser in~\cite{Khaneja2001} and~\cite{PhysRevA.63.032308}. Recall that all maximal Abelian subalgebras share an adjoint orbit, namely $\mathfrak{m}$ itself, and that one may, as a result, switch between them with relative ease.

\subparagraph{Definition:} Nakajima, Kawano and Sekigawa choose to define 
\be \mathfrak{h}={\rm span}_\mathbb{R}\{|j\rangle\langle j|\otimes i\sigma_{x}|j=0,...,2^{n-1}-1\}\ee
  
The algorithm of~\cite{nakajima-2005} based upon this choice of $\mathfrak{h}$ corresponds to a decomposition of an $n$-qubit quantum logic circuit into $2^{n-1}-1$ uniformly controlled one qubit elementary rotations, requiring $4^n - 2^{n-1}$ CNOT gates.  

Note that Nakajima, Kawano and Sekigawa  Cartan decompose $SU(2^n)$ yielding 2 elements of $SU(2^{n-1})\oplus SU(2^{n-1})$.  These are then implicitly treated as if they were 4 elements of $SU(2^{n-1})$ with no further special structure, and  precisely the same Cartan decomposition is applied to each of these smaller unitary operators.  This approach is implicitly based on the assumption that the tensor sum of Cartan decompositions is the Cartan decomposition of tensor sums.  This assumption, however, can easily be proven to be false.  Thus, we now set out to find a Cartan decomposition of the Lie algebra $\mathfrak{s[u}(2^{n-1}) \oplus \mathfrak{u}(2^{n-1})]$.  

Consider the basis of $\mathfrak{s[u}(2^{n-1}) \oplus \mathfrak{u}(2^{n-1})]$: ${\rm span}_\mathbb{R}\{A\otimes Z,B\otimes{\bf 1},  iZ^{(n)}|A,B\in\mathfrak{su}(2^{n-1})\}$.  
\subparagraph{Definition:} It is straightforward to confirm that the decomposition 
\begin{eqnarray}
\mathfrak{k'} &=& {\rm span}_\mathbb{R} \{A\otimes{\bf 1}, iZ^{(n)}|A \in \mathfrak{su}(2^{n-1})\} \nonumber\\
\mathfrak{m'} &=& {\rm span}_\mathbb{R} \{A\otimes Z|A \in \mathfrak{su}(2^{n-1})\}\nonumber 
\end{eqnarray}
satisfies the definition of a Cartan decomposition for $\mathfrak{s[u}(2^{n-1}) \oplus \mathfrak{u}(2^{n-1})]$.  Notice that $ Z^{(n)}$ represents a phase and commutes with every element of $\mathfrak{k'}$, indeed it commutes with every element of $\mathfrak{s[u}(2^{n-1}) \oplus \mathfrak{u}(2^{n-1})]$.  We may factor out the $ Z^{(n)}$ component from $\mathfrak{s[u}(2^{n-1}) \oplus \mathfrak{u}(2^{n-1})]$ to get $\mathfrak{su}(2^{n-1})\oplus\mathfrak{su}(2^{n-1})$.  If we define $\widetilde{\mathfrak{k'}}=\mathfrak{k'}\setminus {\rm span}_\mathbb{R}  Z^{(n)}$, then $\mathfrak{su}(2^{n-1})\oplus\mathfrak{su}(2^{n-1}) = \widetilde{\mathfrak{k'}}\oplus \mathfrak{m'}$ is a Cartan decomposition.  
\subparagraph{Definition:} A Cartan involution to separate these subsets is $\theta (m) = X^{(n)} m X^{(n)}$.  Furthermore we find that if we apply this involution to an element of $\mathfrak{s[u}(2^{n-1}) \oplus \mathfrak{u}(2^{n-1})]$ which has \textit{not} had its $ Z^{(n)}$ phase factored out, the phase lands in the $-1$ eigenspace.  We must also choose a Cartan subalgebra in $\mathfrak{m'}$; for simplicity, we choose the set of diagonal elements of $\mathfrak{m'}$: $\mathfrak{h'} ={\rm span}_\mathbb{R} i \{IZZ,ZIZ,ZZZ\}$ in the three qubit case.  

We now compute the Cartan KAK factors of an arbitrary element ($G$) of $S[U(2^{n-1}) \oplus U(2^{n-1})]$.  First we use the method of Theorem 2 to compute the component of $G$ \textit{not} in exp($\widetilde{\mathfrak{k'}}$), i.e. we compute $\tilde{M}^2 = M^2 P^2$ where $M$ is from  $G=KM$ and $P$ is the $ Z^{(n)}$ factor.  Next we diagonalize  $\tilde{M}^2$ - this diagonal matrix is $A^2 P^2$, where $M = LAL^\dagger$ for $A\in \exp(\mathfrak{h'}), \: L\in \exp(\widetilde{\mathfrak{k'}})$.  Finally we take the square root of this diagonal matrix and compute $K$. To be completely explicit, we present here the algorithm.    

\paragraph{1.} Compute $\tilde{M}^2 = M^{2}P^2=\Theta(G^{\dagger})G$ where $\Theta (U)=X^{(n)}UX^{(n)}$ (see Theorem 2).

\paragraph{2.} Compute the eigenvalue decomposition of $\tilde{M}^2$: let $\tilde{M}^2=LD^2 L^{\dagger}$ be the eigenvalue decomposition.  Since $D^2$ is diagonal and unitary it must be an element of the exponentiation of $\mathfrak{h'} \cup {\rm span}_\mathbb{R} Z^{(n)}$ and L must be an element of $\exp(\widetilde{\mathfrak{k'}})$.

\paragraph{3.} Compute $\tilde{A} = D^{1/2} = AP$ where $A\in \exp(\mathfrak{h'})$ and $P$ is the phase term.  Each entry in the diagonal unitary D is of the form $e^{i\theta}$, so we may simply replace each of these entries with $e^{\frac{i\theta}{2}}$ and we have $\tilde{A}$.  Now $\tilde{M} = L\tilde{A}L^{\dagger}$.  

\paragraph{4.} Compute $K = G\tilde{M}^{\dagger}$.  We have $G=P(KLAL^{\dagger})$ where $P$ commutes with all of the other factors and therefore may be placed according to convenience, $K,L \in \exp(\mathfrak{k'})$ and $A \in \exp(\mathfrak{h'})$, that is $K$ and $L$ are general $(n-1)$ qubit operations which leave the low qubit fixed and $A$ is a uniformly controlled $z$-rotation on the low qubit.  

The operations in $\exp(\mathfrak{k'})$ do nothing to the $n^{th}$ qubit and can perform any unitary operation on the remaining $n-1$ qubits, i.e. we can treat them precisely as we would any element of $SU(2^{n-1})$, and we may absorb the diagonal $P$ into $A$ and implement $\tilde{A} = AP$ according to the decomposition offered in~\cite{Synthesis}, which leaves us with a uniformly controlled z-rotation on the low qubit and a diagonal operator acting on the remaining qubits which may simply be absorb into a neighboring $n-1$ qubit operation.  

\begin{figure}
\begin{center}
\includegraphics[width=1.0\textwidth,viewport=-10 170 400 600,clip]{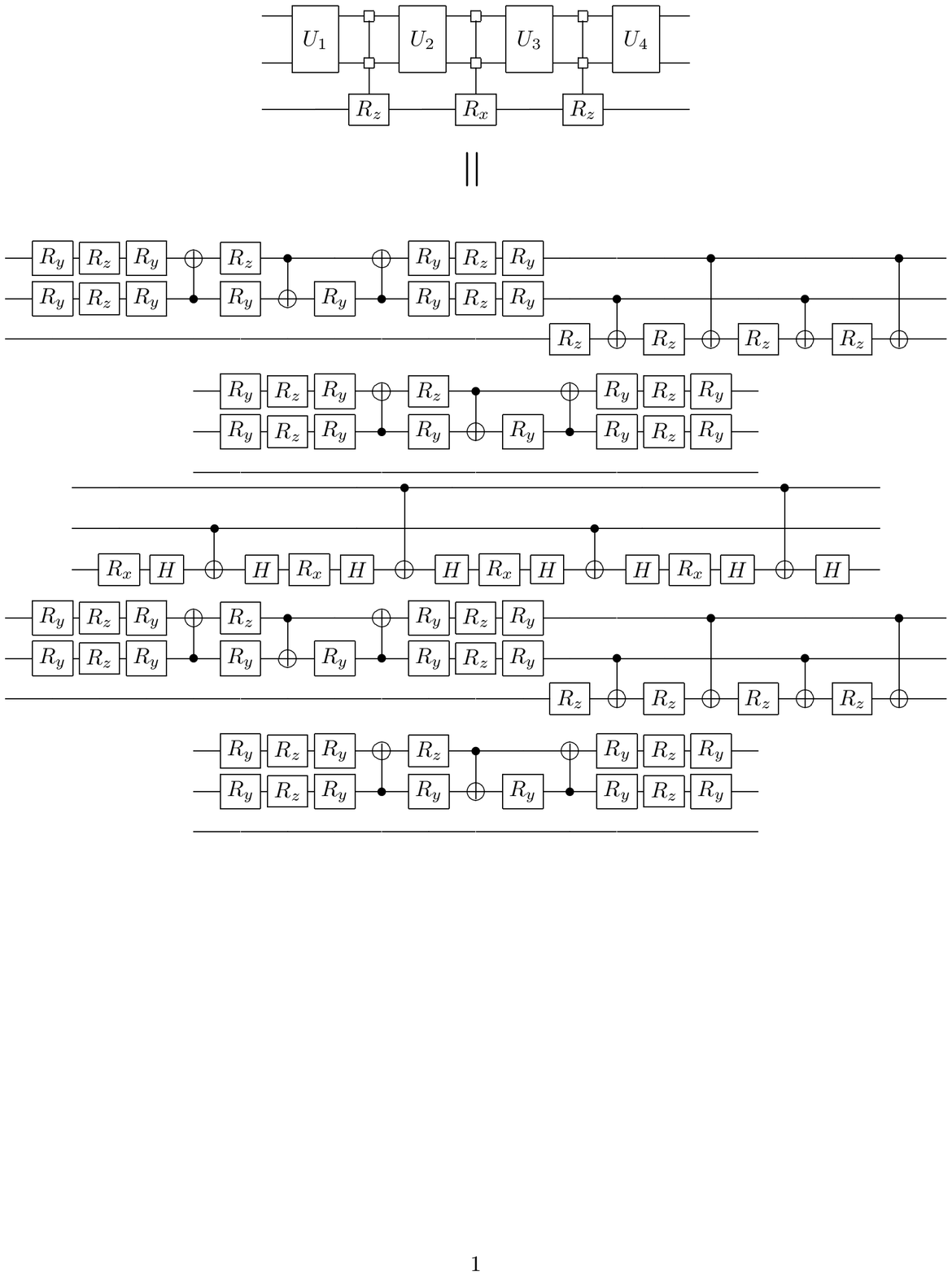}
\caption{The 24 CNOT universal three qubit quantum circuit derived without further simplification from the Cartan decomposition of $\mathfrak{s[u}(2^{n-1}) \oplus \mathfrak{u}(2^{n-1})]$.\label{quack}}
\end{center}
\end{figure}

\begin{figure}
\begin{center}
\includegraphics[width=1.0\textwidth,viewport=0 200 450 400,clip]{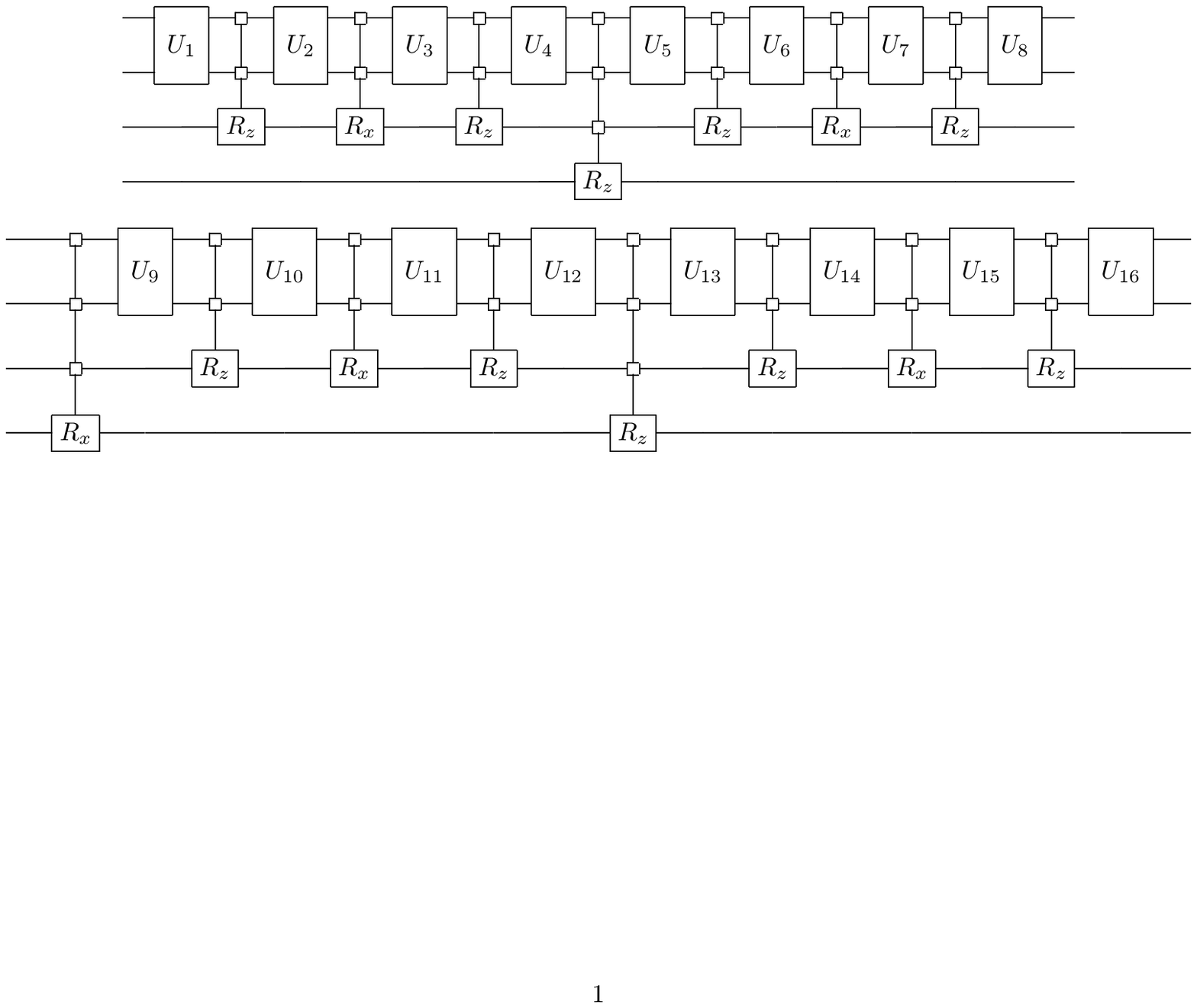}
\caption{\label{brick}A block diagram of the QSD applied to a four qubit operation; notice that it consists of only 3 thrice controlled rotations on the low qubit and 4 general three qubit QSD circuits on the higher qubits.}
\end{center}
\end{figure}

Given an operation on any number of qubits $n$, we apply Nakajima, Kawano and Sekigawa's algorithm to produce 2 elements of $S[U(2^{n-1})\oplus U(2^{n-1})]$, then we apply the algorithm we have just described to these uniformly controlled operations to yield 4 elements of $SU(2^{n-1})$ to which we apply the NKS algorithm, and so on, until we are left with $4^{n-2}$ two qubit operations, to which we apply the \textbf{AI} algorithm described earlier.  This recursive decomposition scheme generates a complete constructive factorization (see Figure \ref{quack} for the three qubit case and Figure~\ref{brick} for an illustration of the recursion applied to four qubits).  Using no further refinements, this algorithm yields a 24 CNOT three qubit gate and has an asymptotic CNOT cost of $\frac{9}{16}4^n - \frac{3}{2}2^n$, an improvement of nearly a factor of two over the standard NKS circuit.  

\section{Conclusions and Future Work}

This scheme of alternating Cartan decompositions of $\mathfrak{su}(2^n)$ with Cartan decompositions of $\mathfrak{s[u}(2^{n-1}) \oplus \mathfrak{u}(2^{n-1})]$ is the best known circuit decomposition paradigm.  This chain of decompositions yields precisely the QSD circuit structure that Shende, Bullock and Markov derived by analogy from the classical Shannon decomposition in~\cite{Synthesis}.  Further slight improvements can be made to the CNOT cost of the tensor sum Cartan circuit by the application of the identities given in Appendix A and Theorem (14) of~\cite{Synthesis}, reducing the overall cost of a three qubit gate to 20 CNOTs, and the asymptotic cost to $\frac{23}{48}4^n-\frac{3}{2}2^n+\frac{4}{3}$,  but the decomposition is still fundamentally the same, and these simplifications can be incorporated into the constructive algorithm presented here with very little effort.  By constructing the QSD from its Lie algebraic roots this work puts the QSD - the best known generic quantum circuit decomposition, less than a factor of two from the highest lower bound - into its proper Lie algebraic context as a series of Cartan decompositions, and provides a new Cartan involution based algorithm to implement the QSD explicitly.  

Another significant advantage of this sort of decomposition, especially in light of the fact that historically few-qubit circuit optimization has at times advanced ahead of asymptotic circuit optimization (cf.~\cite{vatan-2004}), is that any future improvements to few-qubit efficiency can simply be plugged into this algorithm at its lowest level of recursion (where we turn to Vatan and Williams' circuit in this case) and translated instantly into improved asymptotic gate counts. For example, one could use existing methods (e. g. ~\cite{makhlin,SBMsmall}) to test whether a particular two-qubit gate has non-generic structure which means that it requires one or two CNOT gates rather than three. Substantially shorter circuits could be obtained by the application of such methods, and by their extension to three qubit circuits.

\end{document}